\documentclass[twocolumn,aps,prd,showpacs,nofootinbib]{revtex4}

\usepackage{amsmath}
\usepackage{amssymb}
\usepackage{graphicx}

\begin{document}

\title{Extra force in $f(R)$ modified theories of gravity}

\author{Orfeu Bertolami}
\email{orfeu@cosmos.ist.utl.pt}
\affiliation{Instituto Superior T\'ecnico \\
             Departamento de F\'\i sica and Centro de F\'\i sica dos Plasmas,\\
             Av. Rovisco Pais 1, 1049-001 Lisboa, Portugal}

\author{Christian G. B\"ohmer}
\email{christian.boehmer@port.ac.uk}
\affiliation{Institute of Cosmology \& Gravitation,
             University of Portsmouth, Portsmouth PO1 2EG, UK}

\author{Tiberiu Harko}
\email{harko@hkucc.hku.hk}
\affiliation{Department of Physics and Center for Theoretical
             and Computational Physics, The University of Hong Kong,
             Pok Fu Lam Road, Hong Kong}

\author{Francisco S.~N.~Lobo}
\email{francisco.lobo@port.ac.uk}
\affiliation{Institute of Cosmology \& Gravitation,
             University of Portsmouth, Portsmouth PO1 2EG, UK}
\affiliation{Centro de Astronomia e Astrof\'{\i}sica da
Universidade de Lisboa, Campo Grande, Ed. C8 1749-016 Lisboa,
Portugal}

\date{\today}

\begin{abstract}
The equation of motion for massive particles in $f(R)$ modified theories of
gravity is derived. By considering an explicit coupling
between an arbitrary function of the scalar curvature, $R$, and
the Lagrangian density of matter, it is shown that an extra force arises.
This extra force is orthogonal to the four-velocity and the
corresponding acceleration law is obtained in the weak field
limit. Connections with MOND and with the Pioneer anomaly are
further discussed.
\end{abstract}


\maketitle

\section{Introduction}

Higher-order curvature theories of gravity have recently received
a great deal of attention in connection with the possibility of
giving rise to cosmological models in the context of which one can
address the issue of the accelerated expansion of the universe
without the need of {\it ad hoc} scalar fields
\cite{Carroll:2003wy,Capozziello,NojOdint1,NojOdint2,Amendola,BCL}.
These theories involve corrections to the Einstein-Hilbert action
by considering a nonlinear function of the curvature scalar,
$f(R)$. Earlier interest in $f(R)$ theories was motivated by
inflationary scenarios as for instance, in the Starobinsky model,
where $f(R)=R-\Lambda + \alpha R^2$ was considered
\cite{Starobinsky:1980te}. Other motivations include the search
for wormhole-type solutions \cite{B90}. In these studies,
different approaches are used throughout the literature. These
include the metric formalism, where the action is varied with
respect to the metric; the Palatini formalism, where the metric
and the connections are treated as separate variables; and the
metric-affine formalism, which generalizes the Palatini variation,
where the matter part of the action depends and is varied with
respect to the connection \cite{metric-affine}.

Recently, it has been argued that most models proposed so far in
the metric formalism violate weak-field solar system constraints
\cite{solartests}, although viable models do exist
\cite{solartests2}. Furthermore, it has been argued that higher
order gravity may explain the flatness of the rotation curves of
galaxies \cite{CaCaCaTr04,Mbelek,CaCaTr07}. For instance, in the
context of $f(R)=f_0R^n$ theories, the obtained gravitational
potential is shown to differ from the Newtonian one due to the
appearance of a repulsive term that increases with the distance
from the center. The rotation curves of our Galaxy were studied,
and compared with the observed data, so to assess the viability of
these theories and to estimate the typical length scale of the
correction. It was shown at first approximation, where spherically
symmetric and thin disk mass distributions were considered, that a
good agreement with data can be obtained with just the stellar
disk and the interstellar gas.

In this paper we aim to derive the equation of motion for massive
particles in a class of generalized gravitational models in which
the Lagrangian of the gravitational field is an arbitrary function
of the curvature scalar. The study of the equation of motion is of
fundamental importance for the understanding of the structure and
properties of gravitational theories. One of the most effective
ways to test gravitational theories is by matching their
predictions with the motion of real objects. For this purpose, we
point out that the covariant conservation equation for a symmetric
energy-momentum tensor, corresponding to matter is not, in
general, conformally invariant \cite{Wald}. One is led to relax
the covariant conservation of the matter energy-momentum by
considering a coupling between the matter Lagrangian and an
arbitrary function of the curvature scalar. It is interesting to
note that nonlinear couplings of matter with gravity were analyzed
in the context of the accelerated expansion of the Universe
\cite{Odintsov}, and in the study of the cosmological constant
problem \cite{Lambda}. This is reminiscent of the situation in
scalar-tensor theories of gravity and also arises in string
theory. Non-minimal couplings have also been extensively
considered in the literature, namely, between a scalar field and
matter, including baryons and dark matter
\cite{Amendola2,Billyard:2000bh,Holden:1999hm,Mbelek,BM00,Pavon,Pop}.
These couplings imply the violation of the equivalence principle,
which is highly constrained by solar system experimental tests
\cite{Faraoni,BPT06}. However, it has been recently reported, from
data of the Abell Cluster A586, that interaction of dark matter
and dark energy does imply the violation of the equivalence
principle \cite{BPL07}. Notice that the violation of the
equivalence principle is also found as a low-energy feature of
some compactified version of higher-dimensional theories.

In what follows, by considering a coupling between a function of
the curvature scalar and the matter Lagrangian, we show that in
higher-order curvature theories of gravity the equation of motion
of massive particles is non-geodesic. Thus, as shall be shown in
Section \ref{Sec:II}, the equation describing the trajectory of
the particle has an extra-force term, which is orthogonal to its
four-velocity. The non-geodesic nature of motion is a distinct
feature of these $f(R)$ theories, found also in the context of a
scalar field model with a suitable potential and proposed
\cite{BParamos04} as a solution for the Pioneer anomaly problem
\cite{Anderson}. We shall discuss this question in Section
\ref{Sec:III}.
Furthermore, it will also be addressed how our approach can be
regarded as a covariant realization of a Modified Newtonian
Dynamics (MOND) \cite{Milgrom}, even though free from the
arbitrariness of its phenomenological version and somewhat simpler
than its more recent realization which involves besides gravity, a
vector and two scalar fields, the so-called TeVeS approach
\cite{Bekenstein04}.

\section{The equation of motion in $f(R)$ gravitational theories}
\label{Sec:II}

The action for the modified theories of gravity considered in this
work takes the following form
\begin{equation}
S=\int \left\{\frac{1}{2}f_1(R)+\left[1+\lambda f_2(R)\right]{\cal
L}_{m}\right\} \sqrt{-g}\;d^{4}x~,
\end{equation}
where $f_i(R)$ (with $i=1,2$) are arbitrary functions of the Ricci
scalar $R$ and ${\cal L}_{m}$ is the Lagrangian density corresponding to
matter. Note that the strength of the interaction between $f_2(R)$ and
the matter Lagrangian is characterized by a coupling constant
$\lambda$. Analogous nonlinear gravitational couplings with a
matter Lagrangian were also considered in the context of proposals
to address the cosmic accelerated expansion \cite{Odintsov}, and in the analysis
of the cosmological constant problem \cite{Lambda}.

Varying the action with respect to the metric $g_{\mu \nu }$
yields the field equations, given by
\begin{multline}
F_1(R)R_{\mu \nu }-\frac{1}{2}f_1(R)g_{\mu \nu }-\nabla_\mu
\nabla_\nu \,F_1(R)+g_{\mu\nu}\square F_1(R)\\
=-2\lambda F_2(R){\cal L}_m R_{\mu\nu}+2\lambda(\nabla_\mu
\nabla_\nu-g_{\mu\nu}\square){\cal L}_m F_2(R)\\
\hspace{1cm}+[1+\lambda f_2(R)]T_{\mu \nu }^{(m)}~,
\label{field}
\end{multline}
where we have denoted $F_i(R)=f'_i(R)$, and the prime represents
the derivative with respect to the scalar curvature. The
matter energy-momentum tensor is defined as
\begin{equation}
T_{\mu \nu
}^{(m)}=-\frac{2}{\sqrt{-g}}\frac{\delta(\sqrt{-g}\,{\cal
L}_m)}{\delta(g^{\mu\nu})} ~.
\end{equation}

Now, taking into account the covariant derivative of the field
Eqs. (\ref{field}), the Bianchi identities, $\nabla^\mu
G_{\mu\nu}=0$, and the identity
\begin{equation}
(\square\nabla_\nu -\nabla_\nu\square)F_i=R_{\mu\nu}\,\nabla^\mu F_i ~,
\end{equation}
one finally deduces the relationship
\begin{equation}
\nabla^\mu T_{\mu \nu }^{(m)}=\frac{\lambda F_2}{1+\lambda
f_2}\left[g_{\mu\nu}{\cal L}_m- T_{\mu \nu
}^{(m)}\right]\nabla^\mu R ~. \label{cons1}
\end{equation}
Thus, the coupling between the matter and the higher derivative
curvature terms describes an exchange of energy and momentum
between both. Analogous couplings arise after a conformal
transformation in the context of scalar-tensor theories of
gravity, and also in string theory. In the absence of the
coupling, one verifies the conservation of the energy-momentum
tensor \cite{Koivisto}, which can also be verified from the
diffeomorphism invariance of the matter part of the action
\cite{Faraoni,Carroll,Magnano}. Note that from Eq. (\ref{cons1}),
the conservation of the energy-momentum tensor is also verified if
$f_2(R)$ is a constant or the matter Lagrangian is not an explicit
function of the metric.

In order to test the motion in our model, we consider for the
energy-momentum tensor of matter a perfect fluid
\begin{equation}
T_{\mu \nu }^{(m)}=\left( \epsilon +p\right) u_{\mu }u_{\nu
}-pg_{\mu \nu } ~,
\end{equation}
where $\epsilon$ is the overall energy density and $p$, the
pressure, respectively. The four-velocity, $u_{\mu }$, satisfies
the conditions $u_{\mu }u^{\mu }=1$ and $u^{\mu }u_{\mu ;\nu }=0$.
We also introduce the projection operator $h_{\mu \lambda }=g_{\mu
\lambda }-u_{\mu }u_{\lambda }$ from which one obtains $h_{\mu
\lambda }u^{\mu }=0$.

By contracting Eq. (\ref{cons1}) with the projection operator
$h_{\mu \lambda }$, one deduces the following expression
\begin{multline}
\left( \epsilon +p\right) g_{\mu \lambda }u^{\nu }\nabla_\nu
u^{\mu} -(\nabla_\nu p)(\delta_\lambda^\nu-u^\nu u_\lambda)\\
-\frac{\lambda F_2}{1+\lambda f_2}\left({\cal
L}_m+p\right)(\nabla_\nu R)\,(\delta_\lambda^\nu-u^\nu
u_\lambda)=0 ~.
\end{multline}
Finally, contraction with $g^{\alpha \lambda }$ gives rise to the
equation of motion for a fluid element
\begin{equation}
\frac{Du^{\alpha }}{ds} \equiv \frac{du^{\alpha }}{ds}+\Gamma _{\mu
\nu }^{\alpha }u^{\mu }u^{\nu }=f^{\alpha }~, \label{eq1}
\end{equation}
where we have introduced the space-time connection $\Gamma _{\mu
\nu }^{\alpha }$, which is expressed in terms of the Christoffel
symbols constructed from the metric, and where
\begin{eqnarray}
\label{force}
f^{\alpha }&=&\frac{1}{\epsilon +p}\Bigg[\frac{\lambda
F_2}{1+\lambda f_2}\left({\cal L}_m+p\right)\nabla_\nu
R+\nabla_\nu p \Bigg] h^{\alpha \nu }\,.
\end{eqnarray}

As one can immediately verify, the extra force $f^{\alpha }$ is
orthogonal to the four-velocity of the particle,
\begin{equation}
f^{\alpha }u_{\alpha }=0~,
\end{equation}
which can be seen directly from the properties of the projection
operator. This is consistent with the usual interpretation of the
force, according to which only the component of the four-force
that is orthogonal to the particle's four-velocity can influence
its trajectory.

Notice that massless particles do follow geodesics and therefore for them
$f^{\alpha }=0$.

\section{The acceleration law in $f(R)$ gravity}
\label{Sec:III}

To derive the acceleration law in the $f(R)$ gravity, we start by
assuming that the motion of a particle in a space-time with
metric $g_{\mu \nu }$ is given by Eq. (\ref{eq1}). The presence of
the extra force $ f^{\alpha }$ implies that the motion of the
particle is non-geodesic. For $f^{\alpha } = 0$ we recover the
geodesic equation of motion. The usual gravitational effects, due
to the presence of an arbitrary mass distribution, are assumed to
be contained in the term $a_{N}^{\alpha }=\Gamma _{\mu \nu
}^{\alpha }u^{\mu }u^{\nu }$. In three dimensions and in the
Newtonian limit, Eq. (\ref{eq1}) can be formally represented as a
three-vector equation of the form
\begin{equation}
\vec{a}=\vec{a}_{N}+\vec{f}~,  \label{eq2}
\end{equation}
where $\vec{a}$ is the total acceleration of the particle,
$\vec{a}_{N}$ is the gravitational acceleration and $\vec{f}$ is
the acceleration (per unit mass) due to the presence of the extra
force. If $\vec{f}=0$, the equation of motion is the usual
Newtonian one, $\vec{a}=\vec{a}_{N}$, which for a point-like mass
distribution is given by $\vec{a}=-GM\vec{r}/r^{3}$.

Taking the square of Eq. (\ref{eq2}) one obtains
\begin{equation}
\vec{f}\cdot \vec{a}_{N}=\frac{1}{2}\left(
a^{2}-a_{N}^{2}-f^{2}\right) , \label{eq3}
\end{equation}
where the dot stands for the three-dimensional scalar product.
Equation (\ref {eq3}) can be interpreted as a general relation
which expresses the unknown vector $\vec{a}_{N}$ as a function of
the total acceleration $\vec{a}$, the extra force $\vec{f}$ and
the magnitudes $a^{2}$, $a_{N}^{2}$ and $f^{2}$. From Eq.
(\ref{eq3}) one can express the vector $\vec{a }_{N}$, as one can
easily verify, in the form
\begin{equation}
\vec{a}_{N}=\frac{1}{2}\left( a^{2}-a_{N}^{2}-f^{2}\right)
\frac{\vec{a}}{ \vec{f}\cdot \vec{a}}+\vec{C}\times \vec{f},
\label{eq4}
\end{equation}
where $\vec{C}$ is an arbitrary vector perpendicular to the vector $\vec{f}$.
In the following, we assume for simplicity, that $\vec{C}=0$.

The mathematical consistency of Eq. (\ref{eq4}) requires that
$\vec{f}\cdot \vec{a }\neq 0$, that is, vectors $\vec{f}$ and
$\vec{a}$ cannot be orthogonal to each other. We consider that
both vectors are parallel. Therefore, we
can represent the gravitational acceleration of a particle in
the presence of an extra force as
\begin{equation}
\vec{a}_{N}=\frac{1}{2}\left( a^{2}-a_{N}^{2}-f^{2}\right) \frac{\vec{a}}{fa}~.
\label{eq5}
\end{equation}

In the limit of very small gravitational accelerations $a_{N}\ll
a$, we obtain the relation
\begin{equation}
\vec{a}_{N}\approx \frac{1}{2}a\left( 1-\frac{f^{2}}{a^{2}}\right) \frac{1}{f}\vec{a}~.
\label{eq6}
\end{equation}
If one denotes
\begin{equation}
\frac{1}{a_{E}} \equiv \frac{1}{2f}\left( 1-\frac{f^{2}}{a^{2}}\right)~,
\label{def}
\end{equation}
then Eq. (\ref{eq6}) can be immediately written as
\begin{equation}
\vec{a}_{N}\approx \frac{a}{a_{E}}\vec{a}~,
\end{equation}
which has a striking resemblance with the equation put forward
phenomenologically in the so-called MOND approach \cite{Milgrom}.
It then follows that $a\approx \sqrt{a_{E}a_{N}}$, and since $
a_{N}=GM/r^{2}$, then $a\approx \sqrt{a_{E}GM}/r=v_{tg}^{2}/r$,
where $ v_{tg}$ is the rotation velocity of the particle
under the influence of a central force. Therefore, it follows that
$v_{tg}^{2}\rightarrow v_{\infty }^{2}=\sqrt{a_{E}GM}$, from which
arises the Tully-Fisher relation $L \sim v_{\infty }^{4}$ as
$v_{\infty }^{4}=a_{E}GM$, where $L$ is the luminosity that is
assumed to be proportional to the mass \cite{Milgrom}.

Notice however, that in the framework of $f(R)$ gravity, $a_{E}$
is not a universal constant as it depends on local curvature
features. This might explain why it is somewhat difficult to match
the whole galactic phenomenology within the framework of MOND (see
e.g. \cite{BParamos06} for a critical assessment). Nevertheless,
this feature of our model opens up quite interesting
possibilities as we will see next.

Indeed, in general, the definition of $a_{E}$, Eq. (\ref{def}),
allows one to formally represent the extra force as a function of
$a$ and $a_{E}$, that is
\begin{equation}
\frac{f}{a_{E}}=-\left(\frac{a}{a_{E}}\right)^{2}\pm \left(
\frac{a}{a_{E}}\right) \sqrt{1+\left(\frac{a}{a_{E}}\right)^{2}}~.
\end{equation}
Hence, through Eq. (\ref{def}), Eq. (\ref{eq5}) can be rewritten as
\begin{equation}
\vec{a}_{N}=\frac{a}{a_{E}}\left[ h\left( \frac{a}{a_{E}}\right)
\left( \frac{a_{N}}{a}\right) ^{2}+1\right] \vec{a}~,
\end{equation}
where
\begin{equation}
h\left( \frac{a}{a_{E}}\right) \equiv \frac{1}{2}\left(
\frac{a}{a_{E}}\right) ^{-1}\left( \frac{a}{a_{E}}\mp
\sqrt{1+\frac{a^{2}}{a_{E}^{2}}}\right) ^{-1} ~.
\end{equation}
Upon substitution into Eq. (\ref{def}), we verify $a^{2}=v_{\infty
}^{4}/r^{2}=a_{E}GM/r^{2}$, which yields for $a_{E}$:
\begin{equation}
a_{E} =\frac{f^{2}r^{2}}{GM}+2f ~.
\end{equation}
Suppose now that $f\sim GM\alpha /r$, where $\alpha $ is a
constant, then in the large $r$ limit, when $f\rightarrow 0$,
$a_{E}\approx \alpha ^{2}$ is a constant, whose numerical value is
determined by the physical properties of the extra force.

Given the environmental nature of this extra force, only
phenomenology can guide us in its identification. In the galactic
context, it seems natural to identify $a_E$ with the $a_0 =
10^{-10}~{\rm m/s^2}$, the threshold acceleration of MOND. On the
other hand, in the solar system and its neighborhood, the Pioneer
anomaly, whether a real effect unrelated with systematic effects
(see e.g. \cite{BParamos07} and references therein), suggests that
$a_E = a_{\rm Pio} =(8.5 \pm 1.3) \times 10^{-10}~{\rm m/s^2}$.
Interestingly, our approach allows for a unified explanation for
these two problems and can account for the fact that the
characteristic acceleration of each class of observations is
somewhat different.

\section{Discussion and Conclusions}

In this work we have studied a class of generalized gravitational
models, in which the Lagrangian density of the gravitational
sector is an arbitrary function of the scalar curvature and an
explicit coupling between the scalar curvature term and the matter
Lagrangian density. Interestingly, we have found that the equation
of motion of massive particles is non-geodesic. Therefore, the
equation describing the trajectory of particles exhibits a term
representing an extra force, which is orthogonal to its
four-velocity. We have also shown that our models have similar
features with the phenomenological approach of MOND, providing an
alternative formulation to this proposal without the need of
introducing, such as in the TeVeS proposal, extra fields besides
the metric and the scalar curvature, which now plays the role of
an additional scalar field. Furthermore, we have shown that the
extra force is consistent with the so-called Pioneer anomaly. A
distinct feature of the formalism outlined in this work is that it
allows to establish a connection between the problem of the
rotation curve of galaxies, via a solution somewhat similar to the
one put forward in the context of MOND, and the Pioneer anomaly,
even though the characteristic acceleration of these two classes
of observation is somewhat different, about $10^{-10}~{\rm
m/s^2}$. Certainly, a more detailed study of the solar system
implications of our model via the parametrized post-Newtonian
analysis remains still to be performed, but it will be considered
elsewhere.

\acknowledgments

The authors thank Valerio Faraoni, Roy Maartens, Jorge
P\'aramos and David Wands for helpful discussions. The work of OB
is partially supported by the Programa Dinamizador de Ci\^{e}ncia
e Tecnologias do Espa\c{c}o of the FCT--Portugal, under the
project PDCTE/FNU/50415/2003. The work of CGB was supported by
research grant BO 2530/1-1 of the German Research Foundation
(DFG). TH is supported by the RGC grant No.~7027/06P of the
government of the Hong Kong SAR. FSNL was funded by
Funda\c{c}\~{a}o para a Ci\^{e}ncia e a Tecnologia (FCT)--Portugal
through the grant SFRH/BPD/26269/2006.

\end{document}